# Experimental test of q-rules

## Richard A. Mould[1]

### Abstract

*An experiment is described that empirically distinguishes the previously proposed q-rules governing the collapse of a wave function, and contrasts it with the conventional idea of a collapse as well as the current leading theory of collapse advanced by Ghirardi and Pearle.*

## Introduction

The most highly developed theory of quantum mechanical state reduction is the GRW/CSL theory of and Pearle [1, 2]. According to that theory, elementary particles undergo a spontaneous collapse that spreads to the macroscopic level through correlations. The rate of collapse is governed by a small hypothetical constant $\lambda$ that has not as yet been observed.

Collett and Pearle proposed an experiment intended to establish an empirical basis of that theory [3]. A micro-disk of aluminum or gold is suspended in a Paul trap at 4.2º K and very low pressure ($5 \cdot 10^{-17}$ Torr), and the disk's angular diffusion rate is observed. The disk with a radius of 200 nm and a thickness of 50 nm is held vertically with its normal lying along the horizontal, while laser photons are directed horizontally toward it. The angular deflection of the photons is therefore a measure of the disk's angular diffusion. The claimed spontaneous reduction of the angular state can then be observed, so the reduction rate λ can be measured. The measurement must take place during a time *between* collisions with atmospheric molecules.

[1] Department of Physics and Astronomy, State University of New York, Stony Brook, New York 11795-3800. http://ms.cc.sunysb.edu/~rmould

If this experiment produces the expected results we would have to conclude that the q-rule theory proposed by the author is incorrect, for these rules posit no constant λ. However, if the experiment does not confirm the diffusion predicted by GRW/CSL, then other alternatives such as the q-rule proposal will remain on the table.

## Collision reduction with sphere

It is simpler to visualize the author's proposed collapse mechanism with a small sphere. Consider a sphere of radius $r_0 \approx 10^{-5}$ cm that is solid aluminum or gold. Imagine that it has expanded to five times that radius as a result of the uncertainty of its momentum. This is shown in Fig. 1a where a number of small dashed spheres representing the minimum volume sphere are circumscribed by a large dashed sphere representing its uncertainty of the sphere's position. An incoming molecule shown as a black dot in Fig. 1b penetrates the extended radius, engaging the sphere at various possible locations in this sphere of uncertain locations. The first encounter shown in Fig. 1b is a simple scattering of the incoming molecule at one of the possible sphere locations. The second possibility is a faux collision as described in the Appendix of a previous paper [4]. Neither one results in a collapse of the wave. The third encounter in Fig. 1b represents a collision with a resulting collapse of the wave a described by the q-rules in Ref. 4. Only three encounters are shown in the figure although there will be a continuum of possible collisions before there is a stochastic hit on one of them. There are a continuum of possible locations of the sphere inside its extended volume.

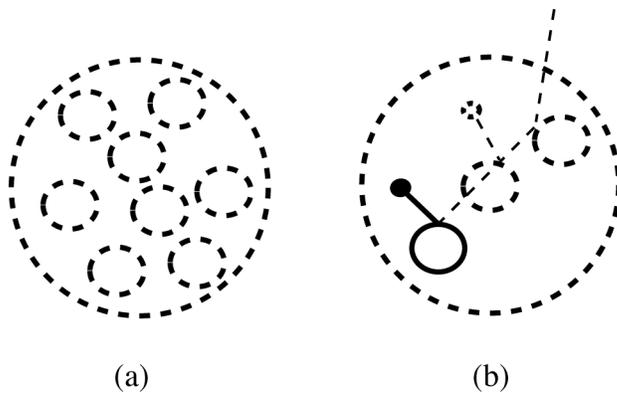

(a)　　　　　　(b)

Figure 1: Uncertainty of position of sphere

If the collisions in Fig. 1b are all continuous like a Compton scattering, there will be no collapse of the wave. For a diatomic molecule at 4.2°K these collisions will no doubt cause many jumps to higher of lower rotational levels, but these alone will not qualify as wave collapses. For a collapse to occur there must be a quantum jump in which a *new particle is created or an old one is annihilated* (Ref. 4). An allowed process is a collision in which the molecule falls to a lower rotational orbit while emitting a photon. The assumption is that of the many collision that occur inside the extended spherical volume, one of them will create a new photon in this way, and that this will satisfying the requirement in Ref. 4. At that point a collapse will localize the sphere and its recoiling molecule as required by to the q-rules, where the sphere now has its minimum volume consistent with its uncertainty of momentum.

## The experiment

The proposed experiment involves a disk rather than a sphere. The reduction principle is the same in both, but a disk has a measurable angular displacement and diffusion rate. According to the q-rule theory, state reductions of the disk will occur only in connection with molecular collisions with the disk, so an observation of a collapse must cover the time before and after a collision in order to confirm the predictions of the theory. The assumption is that between collisions the angular uncertainty $\Delta\phi$ will become much larger than the initial uncertainty $\Delta\phi_0$ (its value right after a collision) because of the initial uncertainty $\Delta L_0$ in angular momentum; and furthermore, that a collapse will reduce the angle to the smallest value $\Delta\phi$ consistent with $\Delta L$ at that time. It will be difficult to measure the state reduction following a collision because of the disruptive influence of the collision;

but assuming that this difficulty can be overcome, a collision reduction will provide a unique test of the proposed q-rule theory inasmuch as no other foundation theory shows that kind of dependence.

Following the process described in the Appendix of Ref. 4, the q-rule equation after the interaction is given by

$$\Psi(t \geq t_0) = sm(t) + \int_0^{a(t-t_0)} \underline{s}\,'m'(t,\tau)d\tau$$

where $s$ is the initial sphere and $m$ is the initial incoming molecule.  The sphere after collision is given by $s'$ and the molecule is given by $m'$.  Each differential contribution to the integrand is a ready component describing what is called a faux collision, but only the first one is a launch component as explained in Ref. 4.  This equation can therefore be written

$$\Psi(t \geq t_0) = sm(t) + \underline{s}\,'m'(t,\tau_0)d\tau + ...$$

where $\tau_0 = 0$.  Again following the Appendix of Ref. 4, a stochastic hit at time $t_{sc}$ yields

$$\Psi(t = t_{sc} > t_0) = s'm'(t_{sc}, \tau_0)$$

The collision dependence of this reduction presents an opportunity for *an* experimental test of the q-rule theory.

I regard the predictions of the q-rules as being well substantiated.  This is based on the exhaustive examples of their application in Ref. 4 and other papers that seem to me 'correct' beyond doubt [5, 6].  This is why I include them with the dynamic principle as part of the mechanics of quantum mechanics.  It is nonetheless possible that they are not fundamental in the same way that the laws of spectroscopy are empirically correct but not fundamental.  It is possible that these rules might be integrated into the dynamic principle in the same way that Ghirardi and Pearle have included a stochastic term into the Hamiltonian.  An integrated theory of that kind seems to me desirable, but to be correct it would have to predict the same experimental results as the q-rules.  The theory of Ghirardi and Pearle does not do that.  It predicts a spontaneous collapse between collisions rather one that occurs *at* a collision with an associated photon emission.